\documentclass[twocolumn, twocolappendix]{aastex631}

\usepackage{epstopdf}
\usepackage{amsmath}
\usepackage{graphicx}  
\usepackage{layouts}

\DeclareMathOperator{\sign}{sign}

\AppendGraphicsExtensions{.gif}

\newcommand\review[1]{#1}

\begin{document}

\title{A Melody in the Noise: Modeling Echoes of the Crab Nebula}

\shorttitle{A Melody to the Noise}
\correspondingauthor{Thierry Serafin Nadeau}
\email{serafinnadeau@astro.utoronto.ca}

\shortauthors{Serafin Nadeau \& van Kerkwijk}

\author[0009-0005-6505-3773]{Thierry Serafin Nadeau}
\affil{David A. Dunlap Department of Astronomy and Astrophysics \\
50 St. George Street \\
Toronto, ON M5S 3H4, Canada}
\affil{Canadian Institute for Theoretical Astrophysics \\
60 St. George Street \\
Toronto, ON M5S 3H8, Canada}

\author[0000-0002-5830-8505]{Marten H. van Kerkwijk}
\affil{David A. Dunlap Department of Astronomy and Astrophysics \\
50 St. George Street \\
Toronto, ON M5S 3H4, Canada}

\begin{abstract}
  Pulses from the Crab pulsar are often followed by ``echoes'', produced by radiation that was deflected by \review{structures} in the Crab nebula and thus traveled via longer paths.
  We describe a simplified but detailed model that treats the \review{structures as cylindrical filaments} of dense, neutral material with a thin ionized skin.
  In this picture, echoes are produced when the line of sight crosses the skin at glancing incidence, which naturally leads to the large electron column density gradients required to get the observed delays even with electron densities comparable to those inferred from optical line emission ratios.
  We compare the properties of the predicted echoes with those of a relatively isolated observed one identified during daily monitoring with CHIME.
  We find that the delays of the simulated echoes follow closely the near quadratic evolution known to be a feature of these echoes, and that, unlike in previous models, we match the characteristic observed asymmetry between incoming and outgoing arcs, with the size of the gap in between a consequence of the skin crossing time.
  However, our model fails to quantitatively reproduce the magnifications of the echoes.
  We believe this likely is because the \review{structures} are not as smooth as envisaged, so that a given echo results from many images.
  Nevertheless, our results \review{confirm that echoes are produced by sheet-like structures seen edge-on and support} the hypothesis that the nebula is filled with small-scale filamentary structures, which may well be substructures of the larger filaments that are seen in optical images.
\end{abstract}

\section{Introduction} \label{sec:intro}

Variations in electron densities of the interstellar medium (ISM) result in corresponding changes in the refractive index of light as it travels from the source to the observer, and hence can lead to lensing of sources, especially at longer radio wavelengths, where the refractive index varies more.
Indeed, these variations are thought to be the cause of changes in quasar flux densities in what are known as extreme scattering events \citep[ESEs;][]{fiedler87, clegg98}.
However, as changes in electron column density cannot be directly measured for quasars due to their continuous emission, ESEs are still poorly understood.
Better understanding such propagation effects is also of interest for the study of fast radio bursts (FRBs) which frequently exhibit structure across time and frequency, and some of which have been localized to highly variable dense and magnetized environments \citep{masui15, michilli18a, thomas22}.

For pulsars, electron density variations are also known to impact the observations, in the form of scintillation due to multipath scattering in the ISM \citep[eg.,][]{scheuer68, rickett90, stinebring01, walker04, brisken10, pen14}.
When the time delays along the multiple paths exceed the duration of the pulses, the pulse profile is broadened by a scattering tail \citep{rankin70, williamson72, williamson74}.
These scattering tails typically have roughly exponential shape, but can also have ``echoes'', discrete features that presumably correspond to discrete scattering structures.

Echoes of pulse emission were first identified in 1997, in a very prominent event in the Crab pulsar \citep{smith00, backer00, lyne01}.
After that, past events were also attributed to echoes (e.g., \citealt{lyne75, vandenberg76, isaacman77}) and less prominent events were identified in later datasets \citep{crossley07, driessen19, rebecca23c}.
Indeed, by stacking of giant pulses in years of monitoring data, we found that echoes are nearly always present \citep{serafinnadeau24}.
Echoes have also been found in a few other sources, e.g., PSR B1508+55 \cite{wucknitz18, bansal20, marthi21} and PSR B2217+47 \citep{michilli18b}, and likely are present whenever the scattering time is long enough to clearly distinguish them.

The Crab is particularly well suited pulsar for studying the structures responsible for echoes because the scattering is not associated with the ISM, but with material from the Crab nebula, which is substantially denser than the ISM, and thus has correspondingly larger breaking indices.
These bend radio waves more easily such that echoes are more common, and are seen even at relatively short wavelengths (echoes have been seen down to 18\, cm; \citealt{rebecca23c}).
Furthermore, the nebular material is seen in optical emission lines and its densities and distances are roughly known, facilitating models.

A benefit of the Crab pulsar beyond the abundance of echoes is that its radio profile is dominated by giant pulses \citep[GPs,][]{staelin68}, extremely bright bursts of emission which unlike regular pulsar emission do not seem to repeat every rotation, but instead appear to occur randomly within a set phase window.
These GPs are very short, lasting only a few micro-seconds, within which they appear to be composed of several nanosecond-long `nanoshots' \citep{hankins03, cordes04}.
Giant pulse emitters are rare, with only a handful of pulsars known to emit them \citep{kuzmin07}, but are ideal for the study of propagation effects because the pulses are so short, which means that the observed scattered profile is a nearly ideal representation of the impulse response function of the intervening medium.
Indeed, in sufficiently narrow frequency bands, giant pulses can be treated as proper impulses, and it has been possible to measure the impulse response function directly, enabling the descattering of nearby emission for some pulsars \citep{main17,mahajan23}.

Unfortunately, for the Crab, coherent measurement of giant pulses is not as useful, as its emission is resolved by the scattering screen \citep{main21, rebecca23a}.
However, by aligning giant pulses before averaging to get a pulse profile, one can avoid the loss in sensitivity and smearing of echoes due to the sporadic nature of giant pulses as well as intrinsic jitter in their arrival times.
Indeed, doing this, we showed that echoes were far more common than implied by the previous rate of detections \citep{serafinnadeau24}.
We also found that the observed echoes are consistent with approaching zero delay at their closest approach to the normal giant pulse emission, which indicates that the structures responsible for producing these events must be highly anisotropic, with typical lengths greater than $\sim\!4{\rm\,au}$, far larger than the typical widths of $\sim\!0.1{\rm\,au}$ inferred from the magnifications.

Given the densities of $\sim\!10^{3}{\rm\,cm^{-3}}$ inferred from optical emission lines \citep{osterbrock57}, we suggested the structures must also be elongated along the line of sight, with depths of $\sim\!5{\rm\,au}$.
Hence, the lensing structures appear sheet-like, as is also inferred for the scattering structures in the ISM that give rise to scintillation \citep{pen14} and possibly to ESEs \citep{romani87, jow24}.

Optical observations show that the Crab nebula is populated by a large quantity of filamentary material \citep{clark83,  graham90, fesen92, martin21}, which is thought to be the product of Rayleigh-Taylor (RT) instabilities formed at the shock between the pulsar wind nebula (PWN) and the as of yet otherwise unseen freely expanding supernova ejecta \citep{hester96, blair97, sankrit98}.

\review{If filamentary structures exist down to the scales probed by scattering,}
edge-on sheets \review{might} arise naturally: as our line of sight crosses a filament, it would necessarily cross also their ionized skins, with particularly large changes in electron column density -- and thus large bending angles -- expected at ingress and egress.
Furthermore, since the filaments are elongated, one expects that most echoes cross zero delay, because when our line of sight passes close enough to a filament to produce an echo, the probability is small that it will not also cross it \citep{serafinnadeau24}.

While the above is plausible, the picture of sheet-like structures differs from what was assumed in previous modeling of Crab echoes.
For instance, \cite{backer00} showed that the 1997 echo event of the Crab could be reproduced by a triangular prism of excess electrons which closely matched the observed dispersion measure (DM) changes, while \cite{graham-smith11}  described the echoes as the result of an ionized filament with a triangular tip, aligned to the pulsar motion and intersecting the line of sight (LoS) to the pulsar.
These models are somewhat ad-hoc and do not naturally explain the preponderance of echoes, most of which cross zero delay.
However, they do reproduce the 1997 echo well.

Here, we use the lensing formalism for folded sheets of \cite{simard18} to see whether the echoes produced by the thin ionized skin of nebular filaments would look like observed ones, comparing in particular to a simple, isolated echo detected in November 2021 during long-term monitoring of the Crab done by \cite{serafinnadeau26}, using the Canadian Hydrogen Intensity Mapping Experiment \citep[CHIME;][]{chime}.
In Section \ref{sec:data}, we briefly describe the data and the basic properties of the echo.
In Section~\ref{sec:model}, we describe the theory behind our model, and in Section \ref{sec:exploration}, we explore the model properties, comparing with the data.
We discuss possible extensions to the model in Section~\ref{sec:extensions}, and ramifications in Section \ref{sec:ramifications}.

\section{An Isolated Echo}\label{sec:data}

For comparisons with our model, we consider a relatively isolated echo observed in November 2021 in daily monitoring of the Crab pulsar with CHIME \citep{chime}.
The Crab monitoring campaign is described in detail in \cite{serafinnadeau26}, but, briefly, we record the beam-formed baseband signal during the Crab's daily $\sim\!15$ minute transit through the main beam, and then search it for giant pulses, keeping the 512 brightest triggers, for each storing $38.4{\rm\,ms}$ of dedispersed  baseband data.
The stored pulses are later aligned, producing daily stacks of pulse intensity, covering the 400--800${\rm\,MHz}$ CHIME band in the standard 1024 channels.

\begin{figure*}[p]
  \centering
  \includegraphics[width=1\textwidth]{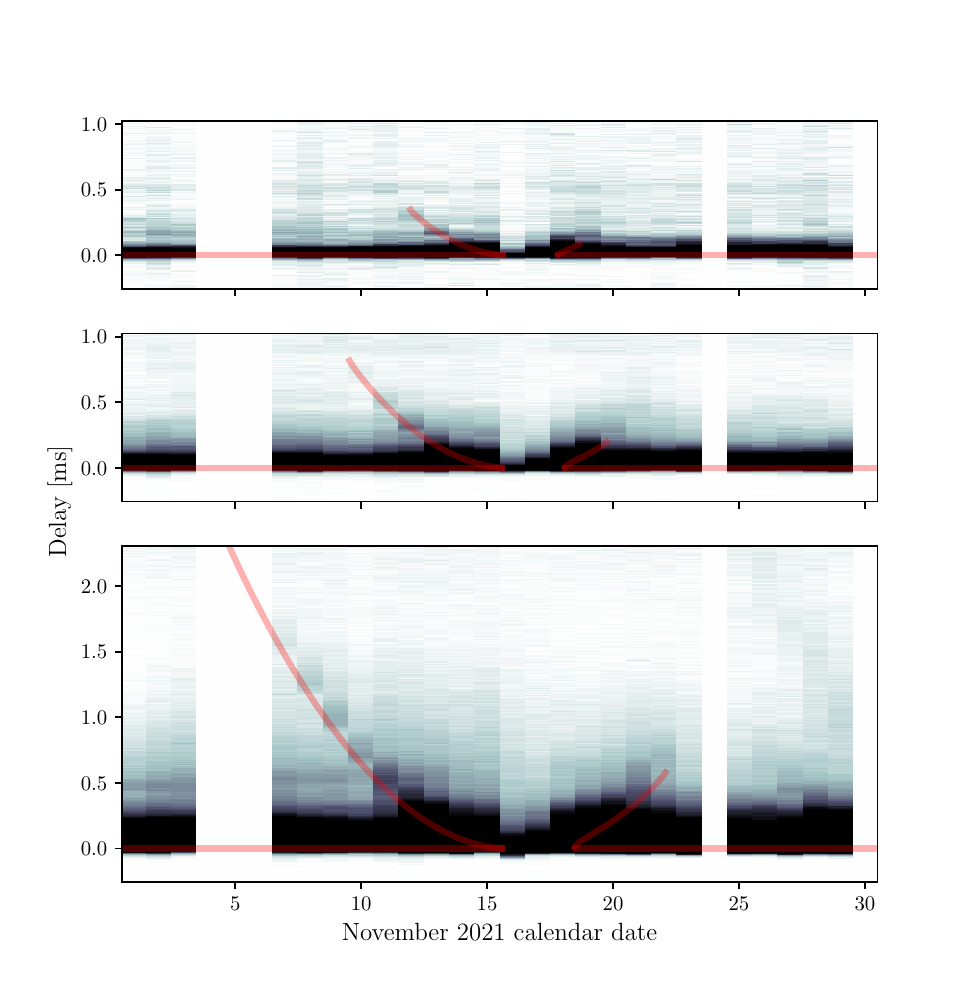}
  \caption{
    Giant pulse profiles in November 2021, averaged over all pulses seen in each daily transit, and over $100{\rm\,MHz}$ bands centered on 750, 600 and $450{\rm\,MHz}$ (top to bottom).
    The positions of the main and echo images expected from the simulations described in Section~\ref{sec:model} are shown in red.
    Note that these do not include any further scattering, i.e., they are expected to trace the bottom of each component.
    \label{fig:nov2021}
  }
\end{figure*}

As can be seen in Figure \ref{fig:nov2021}, the November 2021 event exhibits a pair of asymmetric arcs, becoming visible first at low frequencies on November 7, near a delay of $\sim\!1.5{\rm\,ms}$.
The following days, the delay relative to the main component decreases, until the echo merges with the primary emission on November 15, after which the amplitude is reduced for two days.
A new arc then emerges on November 18, receding from the main component, disappearing from view even at low frequencies on November 22, at a delay of $\sim\!0.75{\rm\,ms}$.

Though the delay between the main component and the echo appears to be largely achromatic, the echo duration depends strongly on the observation frequency, with the additional components only being visible to very short delays near the top of the band.
This is expected, since for a given refractive structure, the maximum bending angle scales as $\nu^{-2}$.

On November 16, as the echo has merged with the main component, we also identified a $0.004\rm{\,pc/cm^3}$ increase in the DM, i.e.,  the lower magnification image was seen through a larger electron column.
Similar jumps in dispersion have been seen before.
For instance, a much larger increase of $0.1{\rm\,pc/cm^3}$ was observed during the 1997 event (e.g., \citealt{backer00}), though that echo was also far more remarkable, both in duration and brightness than the one discussed here, and hence likely associated with a more prominent lensing structure.

\section{Model}\label{sec:model}

\begin{figure}[p]
  \includegraphics[width=\columnwidth]{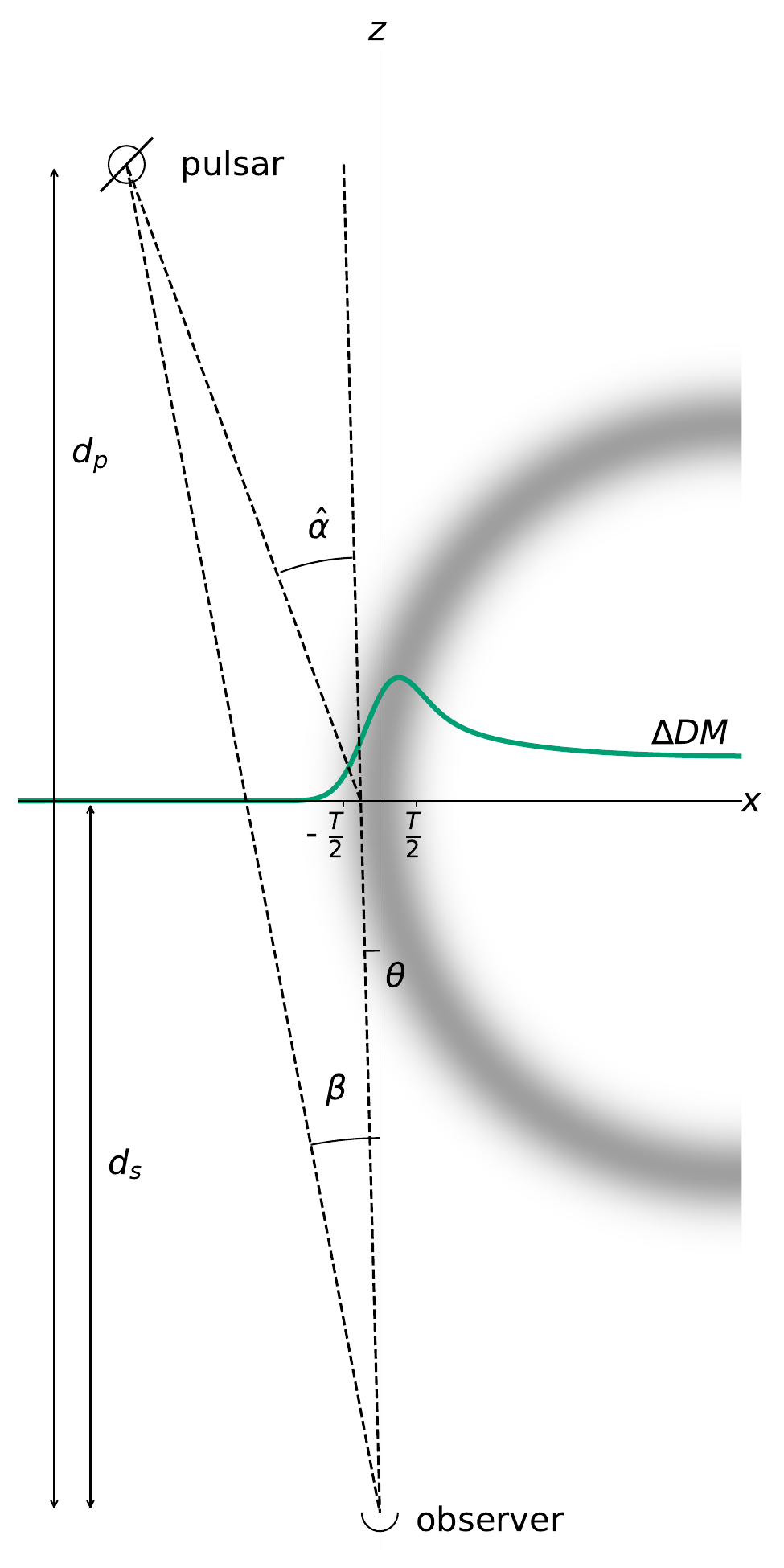}
  \caption{
    Model lensing geometry \review{(adapted from \citealt{simard18})}, assuming a cylindrical filament, perpendicular to the line of sight, with some radius $R$ and a skin with thickness $T=R/10$ with increased electron density (grey shading).
    Lines of sight that cross this filament have increased DM (green line).
    Paths taken by different images are shown, with angles $\beta$ and $\theta$ to the pulsar and lensed image, respectively (both relative to the filament's edge), and bending angle $\hat{\alpha}$.
    Note that this is not to scale: for lensing in the Crab nebula, one has $R\ll d_p-d_s\ll d_s, d_p$, \review{and our model has $T\ll R$}.
    \label{fig:sketch}
  }
\end{figure}

\review{Our model is based on that of \cite{simard18}, in which lensing arises around a fold of a sheet-like structure that is seen nearly edge-on.
  While this model was created for pulsar scintillation in the interstellar medium, similar properties are inferred for the echoes in the Crab: elongation on the sky so that most echoes will approach zero delay, narrow widths to roughly reproduce the magnifications, and elongation along the line of sight to give column density gradients and thus bending angles sufficiently large to produce the observed echo delays \citep{serafinnadeau24}.
Since there is a fold in the sheet, the model is naturally asymmetric, which, less explicitly discussed by \cite{serafinnadeau24}, is consistent with another property of the echoes, viz., that they often are one-sided.}

\review{The echo presented in this paper provides further qualitative confirmation of the basic picture, in two ways.
  First, as the echo delay approaches zero and the pulsar thus moves behind the lensing structure, the intensity decreases, as is expected for an overdense, divergent lens.
  This lasts for about two days, confirming the inference of a thin lens: the pulsar only moves $\sim\!0.1{\rm\,au}$ during that time.
  Indeed, this value is nicely consistent with what was inferred from the magnifications.
  Second, the echo is strongly asymmetric but not quite one-sided, suggesting
different column density gradients on the two sides of the lensing structure.
  In both these properties, it is reminiscent of the striking 1997 event, in which the flux also dimmed (with a larger observed crossing time of $\sim\!10$ days, still implying a small, $\lesssim\!1{\rm\,au}$ width), and the two sides were very asymmetric \citep{backer00, lyne01}.}

\review{Given the inferred sizes of $\sim\!5{\rm\,au}$ and widths of $\sim\!0.1{\rm\,au}$, the sheets are much smaller than can be resolved through current imaging (or simulations), and we are thus left with making informed guesses as to their origin.
The Crab nebula is filled with synchrotron-emitting material that has fine structure \citep{hester95, blair26}, but the typical electron densities in this region are only on the order of $10^{-5} {\rm \, cm}^{-3}$ \citep{shklovsky57}, meaning even in the brighter and thus somewhat denser wisps it is not possible to obtain the electron column density gradients of several $10^4{\rm\,cm^{-2}\,cm^{-1}}$ required to produce the echoes.}

\review{The other primary source of structure in the Nebula are the ejecta filaments, found between $0.5{\rm\,pc}$ from the pulsar and the edge of the nebula \citep{martin21}.
  The ionized surfaces of these filaments, from which the optical emission lines arise, have the highest inferred electron densities in the nebula, up to a few $10^3{\rm\,cm}^{-3}$ (e.g., \citealt{osterbrock57, fesen82, temim24, arias25}).
  If filaments exist down to the scales required for the echoes, it would make them attractive locations for the lensing structures, in particular as they would naturally give rise to large electron density gradients as well, towards the neutral remnant material one one side, and the much less dense pulsar wind nebula on the other.}

\review{Below, to make our model concrete, we will assume sufficiently small filaments exist.}
We sketch the situation we envisage in Figure~\ref{fig:sketch}, where we take the simplest case: a cylindrical filament oriented perpendicular to the line of sight, with an ionized skin with some characteristic width~$T$.
As the pulsar passes behind the filament, the free electron column density will be greatest near the filament's edge, where the path length through its skin is longest.
Near the edge, also the gradients in electron density are largest, causing the strongest lensing, thereby producing echoes.

\review{Before continuing, we would be amiss not to mention that while observations are as yet insensitive to the presence of small filaments, the best angular-resolution observations of the nebula suggest that at least some filaments are resolved (at $\sim\!50{\rm\,mas}$ with HST or JWST, or $\sim\!100{\rm\,au}$ at the Crab, \citealt{sankrit98,loll13,blair26}; and at $\sim\!300{\rm\,mas}$ at low frequencies with LOFAR, \citealt{arias25}), although emission from the lower-ionization state lines, which trace the densest material, appear not to be (they arise in ``very sharp structures'' according to \citealt{sankrit98}).
Furthermore, photo-ionization modeling suggest surface widths much larger than those required for the echoes \citep{pequignot83,sankrit98,priestley17}.
Hence, our hypothesis that the sheets are associated with small filaments may well be wrong.
This, however, does not invalidate the basic constraints our model helps set on the lensing structures.}

\subsection{Modeling the Filaments}\label{sec:dm}

For a cylindrical filament oriented perpendicular to the line of sight, the center of the skin has a circular cross section, with,
\begin{equation}
    \label{eqn:circle}
    x_s^2 + z_s^2 = R^2
\end{equation}
where the coordinates are relative to the center of the filament, with $x$ along the direction of the pulsar proper motion and $z$ along the line of sight (and $y$ is perpendicular to the proper motion and along the filament axis).
Of course, for a general orientation, the cross section will be an ellipse.
However, if the skin is thin, the edge of an ellipse can be well approximated as that of a circle with a local radius of curvature $R_{\rm curv}$.
We will consider the case of an ellipse in more detail in Section~\ref{subsec:ellipse}.

\review{As noted above, photo-ionization models do not produce widths consistent with what we require.
  Generically, there are two competing effects: towards the inside of the filament, the electron density will decrease as material is less ionized, while towards the outside it will decrease because the matter becomes hot and expands.
  For lack of guidance on what the precise electron density profile might be, we} take the (excess) electron density $\Delta n_e(\rho)$ as a function of cylindrical radius $\rho$ to follow a normal distribution around the center of the ionized skin, writing,
\begin{equation}
  \label{eqn:density_profile}
  \Delta n_e(\rho) = n_{e,c}\frac{2}{\sqrt{\pi}} e^{-(2|\rho-R|/T)^2},
\end{equation}
where $n_{e,c}$ and $T$ are the characteristic density and width of the profile, respectively.
This form \review{allows us to explore different profile} shapes with a generalized Gaussian (see Sect.~\ref{subsec:profile}), while keeping \review{the profile smooth and} the electron column density across the skin normalized to $\int_\rho \Delta n_e (\rho)d\rho=n_{e,c} T$.

As the pulsar crosses the structure, the excess electron column density $\Delta {\rm DM}(x)$ will be given by,\footnote{We write the electron column density as ${\rm DM}$ to link it to measurements, but note that the two are not exactly equivalent, both because measurements may be biased by other effect, such as scattering, and because the normal pulsar DM uses a fixed conversion constant, which is not exactly equal to what one would infer from the current values of physical constants.}
\begin{align}
  \label{eqn:dispersion}
  \Delta {\rm DM}(x)
  &= \int_{z=-\infty}^{\infty} \Delta n_e(x, z)\,dz \nonumber \\
  &= 2 n_{e,c}\frac{2}{\sqrt\pi}
      \int_{\rho=|x|}^\infty\frac{e^{-(2|\rho-R|/T)^2}}
                               {\sqrt{1-(x/\rho)^2}}d\rho,
\end{align}
where we implicitly use that the lensing structure is much smaller than the distances to the pulsar and observer (so that $x$ can be assumed to be independent of $z$ along the relevant part of the path), and where the first factor 2 in the second equation accounts for the fact that both sides of the cylinder contribute.

The second form is useful for numerical estimates, as it allows calculating the profile just once, and then summing it multiplied by $[1-(x/\rho)^2]^{-1/2}$.
However, in order to allow for the flexibility to test different skin ionization profiles or variations in filament shapes (e.g. Sect.~\ref{subsec:ellipse} and~\ref{subsec:profile}), we instead numerically evaluate the \review{excess}  column density as the convolution
\begin{equation}
    \label{eqn:convolution}
    \Delta{\rm DM}(x) = \Delta n_e(x+R) * \left( 2 \frac{dl}{dx} \right),
\end{equation}
where $dl$ is the line element defined by $dl^2 = dx^2 + dz^2$.
The above is equivalent to Eq.~\ref{eqn:dispersion} when the line of sight is sufficiently close to the edge of the filament (i.e. $|\rho-R|, |x-R| \ll R$), and this holds true in the lensing region of our simulations, given the observational constraints on the physical parameters (see Sect.~\ref{subsec:implementation}).
To minimize numerical artifacts as a result of this convolution and to ensure sufficient sampling, we use a fine spatial resolution of $2T/1000$, and we evaluate the electron density profile over $3T$ around its center.

It is also useful to make the integral dimensionless.
As shown in Appendix~\ref{sec:appendix}, we find that also allows us to write it in terms of a parabolic cylinder function, $U(a, z)$,
\begin{align}
  \Delta {\rm DM}(x)
  & = 2n_{e, c}\sqrt{RT}\frac{1}{\sqrt\pi}
      \int_{0}^{\infty} \frac{e^{-(\varrho+\xi)^2}}{\sqrt{\varrho}}d\varrho\nonumber\\
  & = 2n_{e, c}\sqrt{RT} \frac{e^{-\frac12\xi^2}}{2^{1/4}}U(0, \sqrt{2}\xi)\nonumber\\
  &\equiv {\rm DM}_{\rm scl} P(\xi),
\label{eqn:scale_and_profile}
\end{align}
where the integral was simplified somewhat using that $T\ll R$ and the dependence on $x$ is implicit via $\xi=-(R/T)(1-(|x|/R)^2)$, with $\xi$ a dimensionless coordinate relative to the center of the skin.
In the final equivalency we separated the scale of the profile, ${\rm DM}_{\rm scl}\equiv2n_{e,c}\sqrt{RT}$ from its shape, $P(\xi)$.

Properties of $P(\xi)$ are discussed in Appendix~\ref{sec:appendix}, but  one has $P(0)=1.023$ for $\xi=0$ ($x=R$), while the maximum is 1.21, at $\xi=-0.54$ ($x=R-0.27T$; see Table~\ref{tab:extrema}), so ${\rm DM}_{\rm scl}$ is indeed the relevant scale near the peak.
Similarly, near the peak, where $\xi\simeq2(|x|-R)/T$ and thus $d\xi/dx\simeq\sign(x)2/T$, one has,
\begin{equation}
    \nabla_x^m {\rm DM} = \sign(x)^m\frac{{\rm DM}_{\rm scl}}{(T/2)^m} \frac{d^mP(\xi)}{d\xi^m},
\end{equation}
where the derivatives of $P(\xi)$ can be expressed in terms of parabolic cylinder functions of different order,
\begin{equation}
    \frac{d^mP(\xi)}{d\xi^m} = (-\sqrt2)^m(e^{-\xi^2/2}/2^{1/4})U(-m, \sqrt{2}\xi).
\end{equation}
Further away from the peak, on the outside of the cylinder, $P(\xi)$ drops exponentially, while on the inside one has $P(\xi)\simeq1/\sqrt{-\xi}$.
Hence, at the center of the cylinder ($x=0$, $\xi=-R/T$), $P(\xi)=\sqrt{T/R}$ and one recovers $\Delta {\rm DM}(0)=2n_{e,c}T$.

In principle, the use of a special function allows quick evaluation, but our code does not use this, since we also allow different density profiles.
But it serves as a test case, and we will use the known derivatives and their roots and extrema for estimates below.

\subsection{Modeling the Images}\label{sec:image_model}

\begin{figure}[p]
  \centering
  \includegraphics[width=\columnwidth]{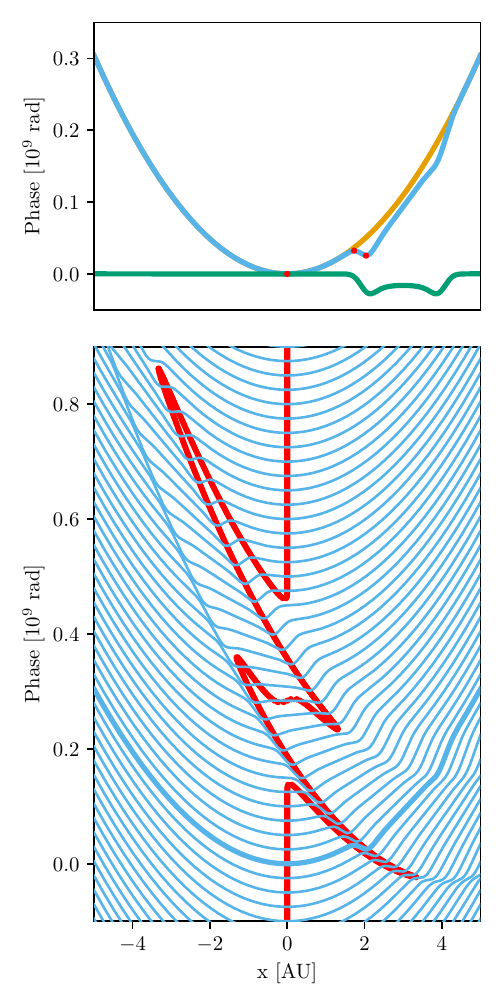}
  \caption{
    Phase changes along different paths, \review{illustrating points of stationary phase, as the pulsar crosses the leading edge of a lens (with parameters chosen for visual clarity; specifically, we used }$\nu = 800{\rm\,MHz}$, $T=0.5{\rm\,au}$, $R=1{\rm\,au}$ and $n_{e,c}=10^5{\rm\,cm^{-3}}$).
    {\em(Top)\/} Situation before ingress.
    The total phase difference relative to the line of sight is shown in blue, with the contributions from the geometric and dispersive components shown in orange and green, respectively.
    Points of stationary phase, where images form, are highlighted as red dots.
    {\em(Bottom)\/} The total phase for \review{multiple screen positions relative to the pulsar, throughout the filament crossing}, as a series of blue curves. \review{The thicker blue line corresponds to the total phase shown in the top panel, while thin blue lines show the total phase at different times, offset along the phase axis to make them visually distinct.}
    Locations of stationary phase are again shown in red.
    \label{fig:phase_screen}}
\end{figure}

\review{As has been done in prior work \citep[eg.,][]{cordes86, clegg98, backer00, simard18}, we} can calculate the effect of our lensing structure on the propagation of light emitted by the pulsar by considering how the phase of emission $\phi$ is altered along different paths (see Fig.~\ref{fig:phase_screen}).
Assuming the eikonal approximation, images will form at points of stationary phase, where
\begin{eqnarray}
    \label{eqn:stationary_phase}
    \nabla_x \phi(x_i) = 0.
\end{eqnarray}

The phase has geometric and dispersive contributions, $\phi = \phi_{\rm geo} + \phi_{\rm disp}$. 
\review{The geometric component can be written as $\phi_{\rm geo} = \omega \tau_{\rm geo}$, with angular frequency $\omega = 2\pi c/\lambda$ and path length delay,
\begin{equation}
  \label{eqn:tau_geo}
  \tau_{\rm geo} = \frac{\left(x-x_p\right)^2}{2 d_{\rm ps} c}.
\end{equation}
Here, $x$ and $x_p$ are the positions of the lens and the pulsar, respectively, $d_{ps}$ is the distance between the pulsar and the lens, and we used that the lens is much closer to the pulsar than the observer and that $d_{ps}\gg x-x_p$ (allowing small-angle approximations).
Rearranging, one recovers the familiar expression in terms of the Fresnel scale $R_F = \sqrt{\lambda d_{\rm ps}}$,
\begin{equation}
    \label{eqn:phi_geo}
    \phi_{\rm geo} = \frac{\pi \left( x - x_p \right)^2}{R_F^2}.
\end{equation}}

\review{Meanwhile, the dispersive component of the phase arises from a difference in phase velocities in and outside of the lens,
\begin{equation}
    \label{eqn:phi_disp_deriv}
    \phi_{\rm disp} = \omega \int\left(\frac{1}{v_\phi}
      - \frac{1}{v_{\phi,o}}\right)dz,
\end{equation}
where $v_{\phi} = c / n$ depends on the index of refraction,
\begin{equation}
  n = \sqrt{1 - \frac{\omega_p^2}{\omega^2}}
  \simeq 1 - \frac{\lambda^2 r_e}{2 \pi} n_e.
\end{equation}
with $\omega_p = \sqrt{4 \pi r_e n_e c^2}$ the characteristic oscillation frequency of a cold plasma and $r_e$ the classical electron radius.
For the approximate equality, we used that $\omega \ll \omega_p$ for typical radio observations.
Combining with Eq.~\ref{eqn:phi_disp_deriv} and~\ref{eqn:dispersion}, we can write $\phi_{\rm disp}$ in terms of the excess electron density or the column density,
\begin{equation}
  \label{eqn:phi_disp}
  \phi_{\rm disp} = -\lambda r_e \int \Delta n_e(x) dz
  = -\lambda r_e \Delta {\rm DM}(x).
\end{equation}
Like the geometric phase, one can write the dispersive phase in terms of the dispersive delay, $\phi_{\rm disp}=-\omega\tau_{\rm disp}$, with the minus sign arising because $\tau_{\rm disp}$ depends on the group velocity, $v_g=nc$, not the phase velocity,
\begin{equation}
    \label{eqn:tau_disp}
    \tau_{\rm disp} = \int\left(\frac{1}{v_g}
      - \frac{1}{v_{g,o}}\right)dz
    = \frac{\lambda^2 r_e}{2 \pi c} \Delta{\rm DM}(x).
\end{equation}
}


Combining the \review{Eqs.~\ref{eqn:stationary_phase}, \ref{eqn:phi_geo} and \ref{eqn:phi_disp}}, one recovers that for the images \review{found at positions} $x_i$ \review{, where the stationary phase condition is satisfied}, the bending angles are determined by the gradient of the column density,
\begin{equation}
  \hat\alpha = \frac{x_i-x_p}{d_{\rm ps}}
  = \frac{\lambda^2r_e}{2\pi} \nabla_x {\rm DM}\big|_{x_i},
  \label{eqn:bending_angle}
\end{equation}
\review{where $\nabla_x {\rm DM} = \nabla_x \left( \Delta {\rm DM} \right)$ because we assume the column density does not vary outside the lens.}
Here, separating the gradient in a dimensionful component and one that depends just on the profile as in Section~\ref{sec:dm}, we find,
\begin{equation}
  \hat\alpha = \frac{\lambda^2r_e}{2\pi} \frac{{\rm DM}_{\rm scl}}{T/2}\left.\frac{dP(\xi)}{d\xi}\right|_{\xi_i} = \frac{T/2}{d_{\rm ps}}fP^\prime(\xi_i),
  \label{eqn:bending_angle2}
\end{equation}
where in the last equality we defined $P^\prime(\xi)\equiv dP(\xi)/d\xi$ and, implicitly, a dimensionless lens strength,
\begin{equation}
  f = \frac{{\rm DM}_{\rm scl}/(T/2)^2}{2\pi/d_{\rm ps}\lambda^2r_e}.
  \label{eqn:lens_strength}
\end{equation}
With these, the image positions can be written as,
\begin{equation}
  \xi_i = \xi_p + fP^\prime(\xi_i)
  \quad\Leftrightarrow\quad
  x_i = x_p + \frac12TfP^\prime(\xi_i).
  \label{eqn:image_position}
\end{equation}

With the image positions $x_i$ known, the \review{corresponding geometric and dispersive} delays \review{$\tau_{\rm geo}(x_i)$ and $\tau_{\rm disp}(x_i)$} can be determined.
For images far from the line of sight, the geometric delay will dominate, and thus depend on the bending angle as $\tau=\frac12\hat\alpha^2 d_{\rm ps}/c$.

For the image magnifications, we use that in geometric optics, the magnification of each lensed image $\mu_i$ can be described in terms of conservation of surface brightness.
In our model the lensing is one-dimensional, along the axis of motion of the system, and one has
\begin{equation}
    \mu_i = \frac{d\theta_i}{d\beta},
\end{equation}
where $\theta_i$ and $\beta$ are the angular positions on the sky of the images and the pulsar (see Fig.~\ref{fig:sketch}).
Since $d_{\rm ps}\ll d_p$, we have \review{$\theta_i = x_i / d_{\rm p}$ and $\beta = x_p / d_{\rm p}$, and it therefore follows that $\mu_i=dx_i/dx_p$}.
Using Eqs.~\ref{eqn:bending_angle} and~\ref{eqn:lens_strength}, we find $x_p=x_i-d_{\rm ps}\hat\alpha$ and
\begin{equation}
  \mu_i = \frac{1}{1-\frac{1}{2\pi}d_{\rm ps}\lambda^2r_e\nabla_x^2 {\rm DM}}
  = \frac{1}{1-fP^{\prime\prime}(\xi_i)},
  \label{eqn:magnification}
\end{equation}
where $P^{\prime\prime}(\xi)\equiv d^2P(\xi)/d\xi^2$.

The above implies magnifications can become infinite, which is obviously not physical.
One can infer what should happen instead by thinking about the magnification in terms of ratios of areas around images in which the radiation is coherent;
for our essentially one-dimensional lens, one has $\mu=(R_i/R_F)^2$ where $R_i$ is, like the Fresnel radius, the range around $x_i$ within which $\phi_{\rm tot}$ is the same within some tolerance.
Expanding Eqs~\ref{eqn:phi_geo} and~\ref{eqn:phi_disp} to second order around an image, one finds $\mu=\nabla_x^2\phi_{\rm geo}/\nabla_x^2\phi_{\rm tot}$, which reproduces Eq.~\ref{eqn:magnification}, but also shows that the infinity would disappear if one expanded to higher order \review{(for a more quantitative analysis, see \citealt{grillo18})}.

In our simulations, we ignore the latter complication, and simply calculate the geometric and dispersive phases and their gradients for each time step and frequency, and then identify \review{the corresponding} image positions \review{$x_i$} by \review{numerically solving for} the roots of the total phase gradient\review{, as per Eq.~\ref{eqn:stationary_phase}}.

\section{Model Properties}\label{sec:exploration}

To see what types of echoes are expected from our simple model, we first estimate model parameters that should roughly match our sample echo\review{, informed by the observational constraints from \cite{serafinnadeau24}}, and then use these to simulate an echo.
Next, we describe qualitatively where the echoes are produced, and then turn to a more quantitative discussion of their properties, focusing first on their delays and durations, and then on their magnifications.
In the process, we compare these predictions qualitatively with the observations, finding good matches for the echo delays and durations, but not for the magnifications.

\begin{figure*}[p]
  \centering
  \includegraphics[width=\linewidth]{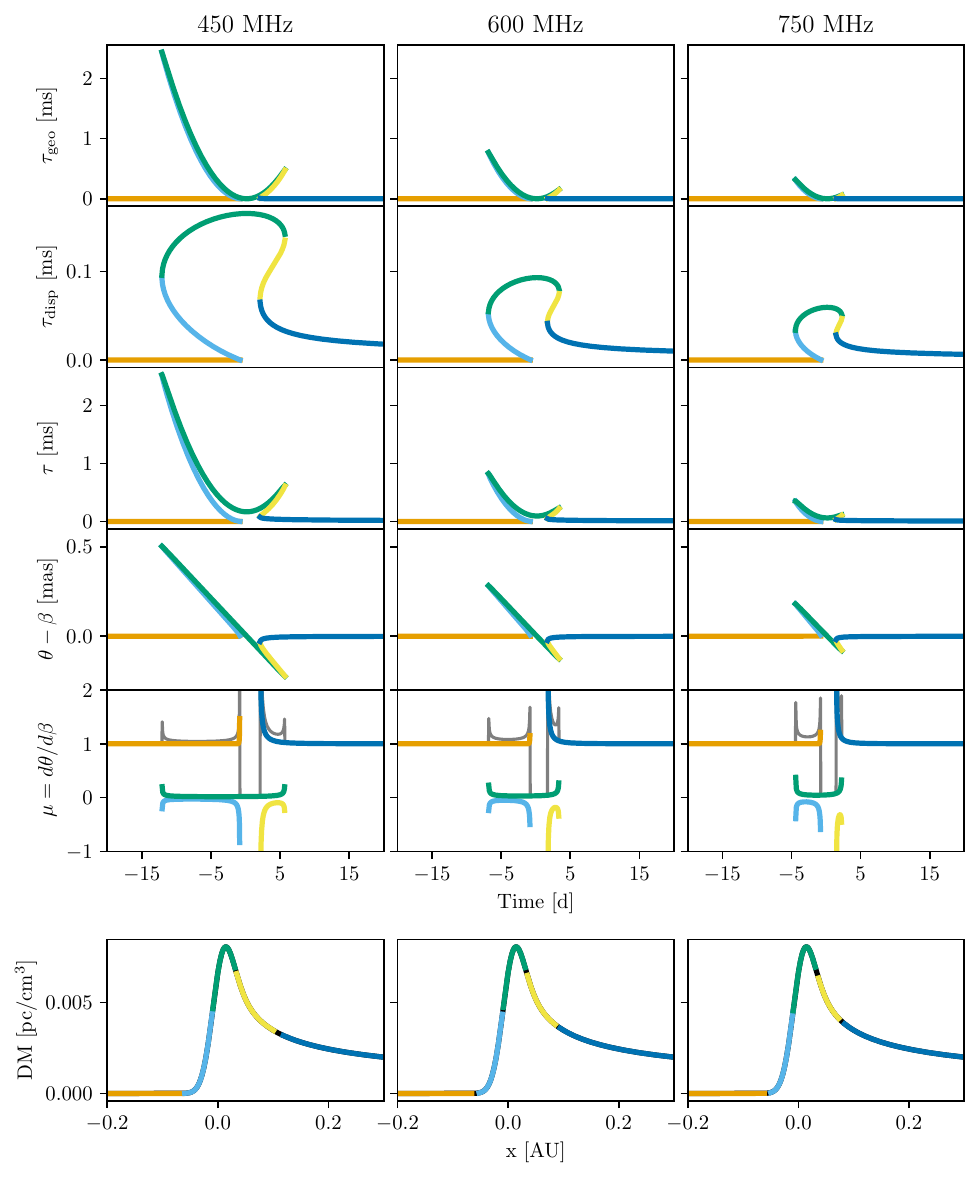}
  \caption{\label{fig:summary}
    Simulated lens properties for three different frequencies (as labeled on top), with parameters chosen to produce delays similar to those seen in the observed echo shown in Fig.~\ref{fig:nov2021}.
    The rows show:
    {\em (1--3)\/}~geometric, dispersive and total delays, as a function of time;
    {\em (4)\/}~offset~$\theta-\beta$ of the lensed image relative to that of the pulsar;
    {\em (5)\/}~image magnifications~$\mu$\review{, with $\Sigma \left|\mu_i\right|$ shown in grey}; and
    {\em (6)\/}~electron column density ${\rm DM}$, as a function of position along the lens.
    In all panels, the images are color coded the same way, to help identify which part of the column density profile they crossed through.
    }

\end{figure*}

\subsection{Model Parameters}\label{subsec:implementation}

\review{As the pulsar moves relative to the screen on the sky, its position changes with time $x_p = v_{\rm ps} t$, where $v_{\rm ps}=v_p-v_s$ is the relative velocity of the pulsar and screen in the $x$ direction.}
The delay at which an echo is seen will generally be dominated by the geometric delay, and thus it will vary quadratically in time, as $\eta\Delta t^2$, with $\eta=v_{\rm ps}^2/2cd_{\rm ps}$.
Assuming the lensing happens in nebular material, one has $0.5 \lesssim d_{\rm ps} \lesssim 1.5{\rm\,pc}$ \citep{martin21}, while the projected velocity is likely of order that of the pulsar, $v_p=120\pm23{\rm\,km/s}$ \citep{kaplan08}, i.e., one expects $\eta\simeq 6{\rm\,\mu s/day^2}(v_{\rm ps}/120{\rm\,km/s})^2/(d_{\rm ps}/1{\rm\,pc})$.

For our sample echo, $\eta\simeq17{\rm\,\mu s/day^2}$, similar to the fastest-evolving echo found by \cite{serafinnadeau24}.
Thus, like in that work, we place the structure at the minimum allowed distance of $0.5{\rm\,pc}$ and use a velocity $v_{\rm ps} = 145{\rm\,km/s}$.
Furthermore, we pick November 16 as the time at which the line of sight to the pulsar crosses the middle of the skin, and let the pulsar move in the direction of increasing~$x$.\footnote{Our simulation code is meant to be general and thus does not assume $d_{\rm ps}\ll d_s$, unlike in the derivation presented in Sect.~\ref{sec:model}.
For completeness, we note that we set $d_p=2{\rm\,kpc}$ for the results presented here; any uncertainties in this have negligible effect.}

For the electron density, we use $n_{e,c} = 1000{\rm\,cm^{-3}}$, matching a typical number from line emission in the nebular filaments, though the observed range varies between 550 and $3500\rm{cm}^{-3}$ \review{in older studies of the Crab nebula} \citep{osterbrock57,fesen82} \review{and the densities inferred from more modern studies also largely fall within this range, in both observation \citep{loh12, temim24, arias25} and simulation \citep{priestley17, das20}}.

\cite{serafinnadeau24} used the \review{observed} echo brightness at maximum delay to estimate a typical width $T \simeq 0.1{\rm\,au}(d_{\rm ps}/1{\rm\,pc})$, \review{and therefore, given the distance $d_{\rm ps} = 0.5 {\rm\, pc}$ inferred from the echo curvature, we set} $T=0.05{\rm\,au}$.
Using the electron density, they then also inferred a typical depth $Z\simeq45T$.
Equating this to the maximum depth one sees when the pulsar crosses the inner radius of the structure, i.e., $Z=2\sqrt{2TR}$, we set $R\simeq10{\rm\,au}$.
For these choices, ${\rm DM}_{\rm scl}=2n_{e,c}\sqrt{TR}=6.86\times10^{-3}{\rm\,pc/cm^3}$ (and the maximum larger by a factor 1.21; see Table~\ref{tab:extrema}).

A summary of the properties of the predicted echo for our default choice of parameters is shown in Figure~\ref{fig:summary}.
The predicted delays are also are overdrawn on our sample echo in Figure~\ref{fig:nov2021}; one sees that these reproduce the observed ones well.

Of particular note is that the model naturally reproduces the asymmetric nature of the echo, where the leading echo arc starts at larger delays and lasts longer than its trailing counterpart.
Similar behavior is clearly visible in the 1997 event presented by \cite{backer00} and \cite{lyne01}, as well as in various echoes shown in \cite{serafinnadeau24}, and so can be considered to be a common feature of Crab echoes.
Note that while this is an expected feature of the model (see below), the parameters we set only used information from the more dominant arm; that the shorter-lasting counterpart is reproduced as well is thus encouraging.

\subsection{Image Trajectories}\label{sec:location}

For the parameters chosen, Figure~\ref{fig:summary} shows that the lens is strong enough over the whole frequency range to produce extra images, which are created and annihilated in pairs.
A pair is first created when maximally bent light from the pulsar is able to reach us.

The stronger of the pair, dubbed the strong echo, is inverted and moves opposite to the pulsar motion on the sky, with its light crossing further and further towards the front face of the lens, and thus should have less and less excess DM.
As the pulsar approaches the ionized skin, this image merges with the main image and both disappear, reappearing later as the pulsar moves past the tangent point.

Meanwhile, the second image of the pair (the weak echo) moves in the same direction as the pulsar, with its ray going through denser and denser parts of the lens, thus having increased DM.
This image does not disappear as the pulsar crosses the skin, instead bridging the gap between the arcs after the main image has disappeared, staying near the peak of the column density.

For a strong event like the one modeled, the duration of the whole echo depends strongly on frequency, \review{with echoes being far more prominent in both delay and duration for lower frequencies.
This is expected given the strong frequency dependence of the bending angle (Eq.~\ref{eqn:bending_angle}), which allows images to form further from the lens as the frequency decreases, and thus at longer delays.
Conversely,} the \review{duration of the} gap during which only the weak echo remains has a much weaker \review{frequency} dependence, though \review{it still is} somewhat longer at lower frequencies (see Fig.~\ref{fig:summary}).
Towards frequencies above the range observed, and for weaker lenses, the gap will get shorter, until it disappears when the lens becomes weak enough that it no longer can form separate images.
Because the lens is not symmetric, this will go through an intermediate case, where the arc arising from the outer edge of the skin still exists, but that of the inner edge, which has less steep column density gradients, has disappeared.

\subsection{Duration and Delays}
\label{sec:delays}

The asymmetry of the echo follows naturally from the leading side having the steeper electron density gradient, and thus has a larger maximum bending angle.
To estimate the maximum bending angles and thus the delays and durations, we first note that since images are produced at points of stationary phase $\nabla_x \phi_{\rm tot} = 0$, they must appear and disappear in pairs, whenever $\nabla_x^2 \phi_{\rm tot} = 0$, or, given Eqs.~\ref{eqn:phi_geo} and~\ref{eqn:phi_disp} for the geometric and dispersive contributions to the phase, when,
\begin{equation}
  \nabla_x^2 {\rm DM} = \frac{2 \pi}{\lambda^2 d_{\rm ps} r_e }
  \quad\Leftrightarrow\quad
  P^{\prime\prime}(\xi) = 1/f,
  \label{eqn:echo_condition}
\end{equation}
where we wrote the second relation in terms of lens strength $f$ and second derivative of the profile $P^{\prime\prime}(\xi)$ using Eqs~\ref{eqn:lens_strength} and~\ref{eqn:magnification}.

Of course, for images to appear, $\nabla_x^2 {\rm DM}$ first has to be large enough.
Inserting numbers for our model, we find $f\simeq26$ for $\lambda=0.5{\rm\,m}$.
Since $P^{\prime\prime}(\xi)$ has positive extrema of $(0.86, 0.27)$ (see Table~\ref{tab:extrema} in App.~\ref{sec:appendix}), separate images are indeed expected for both the incoming and outgoing echo.
Echoes would only disappear at much shorter wavelengths, $\lambda\simeq10{\rm\,cm}$.

Indeed, because the lens is so strong, one can estimate the maximum DM gradient by evaluating it at the positions where $\nabla_x^2 {\rm DM}=0$ (this will be quite accurate, since $\nabla_x {\rm DM}$ necessarily changes only slowly near these points).
Consulting Table~\ref{tab:extrema}, we find extrema  at $\xi_{\pm}=(0.39, -1.23)$ of $P^\prime_{\pm}=(-0.90, 0.41)$.
Using Eqs~\ref{eqn:bending_angle}, \ref{eqn:bending_angle2}, \ref{eqn:image_position} and~\ref{eqn:tau_disp}, this corresponds to bending angles $\hat\alpha_\pm=f(T/2d_{\rm ps})P^\prime_{\pm}=(-1.2, 0.5){\rm\,arcsec}$, delays $\tau_\pm=\frac12\hat\alpha_\pm^2d_{\rm ps}/c=(0.84, 0.18){\rm\,ms}$, pulsar positions $x_{p,\pm}=\frac12T(\xi_{\pm}-fP^\prime_\pm)=(0.24, -0.58){\rm\,au}$, and appearance and disappearance times $x_{p,\pm}/v_{\rm ps}=(-6.9, 2.9){\rm\,d}$, all consistent with our numerical simulations (see Fig.~\ref{fig:summary}).
The properties also scale with wavelength as expected, with bending angles and times scaling with $\lambda^2$, and the delays with $\lambda^4$.

Note that the ratios of the above properties of the points of maximum delay are fixed by the profile, with no free parameters.
Hence, that the predicted asymmetry seems a bit more pronounced than that which is observed, implies that the true column density profile may be more symmetric than modeled, with more similar gradients on each side.
We discuss possible explanations for this difference as extensions to the model in Sections \ref{subsec:profile} and \ref{sec:noisy_skin}.

As can be seen in Figure~\ref{fig:summary}, the echo delays vary quadratically with time.
This is expected since the simulated dispersive component of the delays is small relative to the geometric delay for the stronger images at all times, and becomes important only for the weaker image near when the line of sight passes through the thickest column of material (see rows 1--3 of Figure \ref{fig:summary}).
Indeed, in the context of our model, this has to be the case for any strong lens with $f\gg1$: the ratio of maximum geometric and dispersive delays is
\begin{equation}
    \frac{d_{\rm ps}\hat\alpha_{\rm max}^2 / 2}{\lambda^2r_e\Delta {\rm DM}_{\rm max} / 2\pi} = \frac{f}{2} \frac{P^{\prime2}_{\rm max}}{P_{\rm max}} \simeq \frac{1}{3}f,
\end{equation}
where the approximate equality uses values from Table~\ref{tab:extrema} and thus is specific to the assumption of a Gaussian density distribution.
Nevertheless, one can conceive of an echo where dispersive delays play a larger role than is observed here: e.g., one could increase the width $T$ while also increasing $\Delta n_e\sqrt{R}$ to keep the gradient $\nabla_x {\rm DM}\propto n_{e,c}\sqrt{R/T}$ roughly constant but let $\Delta {\rm DM}\propto \Delta n_e\sqrt{RT}$ increase.

We now turn to the locations where the images appear and disappear close to the lens-crossing.
As discussed in Section~\ref{sec:location} above, these images are produced far out on the profile, and are not associated with extrema like the places where the echo first appears, at large delay.
We can estimate their properties using the asymptotic expansions given in Appendix~\ref{sec:appendix}.
Given that $P^{\prime\prime}=1/f\simeq0.04$, we infer $\xi_{\pm}=(-2.15, 3.23)$ and $P^{\prime}_{\pm}\simeq(-0.019, 0.09)$, the latter corresponding to bending angles of $(-0.02, 0.11){\rm\,arcsec}$.
Thus, the direct image of the pulsar disappears and appears again at $\xi_{p,\pm}=(-2.6, 5.5)$, corresponding to times of $(-0.8, 1.6){\rm\,d}$ (at $\lambda=0.5{\rm\,m}$).
This again matches well with Fig.~\ref{fig:summary}.

\subsection{Magnifications}\label{sec:magnification}

\begin{figure}
  \centering
  \includegraphics[width=\columnwidth]{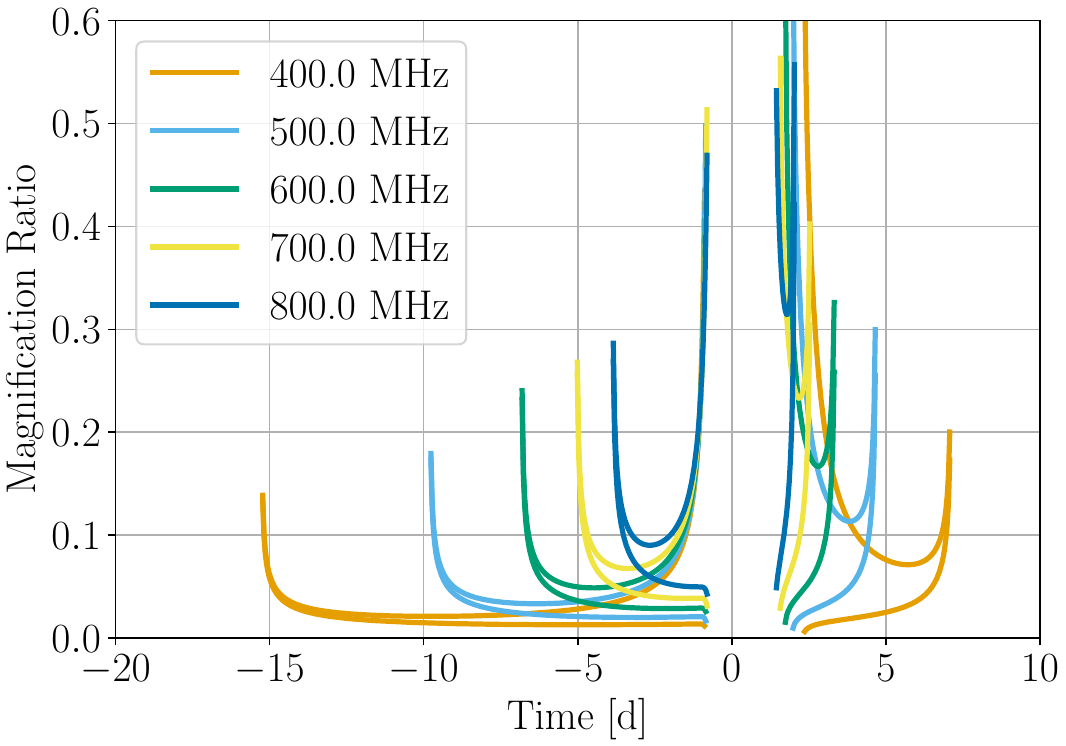}
  \caption{
    Ratio of the magnification of echo images relative to the main component images for a range of frequencies.
    \label{fig:magnification}
  }
\end{figure}

The behavior of the magnification in our simulations can be seen in the fourth row of Figure \ref{fig:summary}.
By far the most striking feature occurs when the images are either produced or disappear.
At these points images are produced in pairs or merge, therefore becoming degenerate in the geometric optics regime, forming caustics where the estimate of the magnification becomes infinite (see Eq.~\ref{eqn:magnification}).
As discussed in Section~\ref{sec:image_model}, those infinities are not physical, but nevertheless a clear prediction is that the brightness should go up near the caustics.
Another clear prediction is that in the inner region, the source should be much less bright than outside, since only a faint image remains.

In between the caustics, the simulated echo magnifications vary smoothly, as can be seen more clearly in Figure~\ref{fig:magnification}, which shows the ratio of the magnification of the two echo images to that of the main image.
Two features stand out.
The first is that both images in the shorter arc are brighter than the corresponding ones in the longer arc.
Qualitatively, this is because those images move more rapidly along the tail of the plasma lens, as can be seen when comparing the position of the echo images, shown on the bottom panel of Figure \ref{fig:summary}, to the echo duration\review{, and so their magnification $\mu_i = dx_i/dx_p$ is enhanced}.
Second, the magnification increases with frequency, where the echoes do not last as long.
Again qualitatively, this results from the duration of the echo shrinking faster with increasing frequency than the region of the lens over which it appears.

The latter suggests there is an optimal range of frequencies over which to observe echoes, where the frequency is low enough that echoes can form and last long enough to be identified, yet high enough that the echoes are bright enough to be seen above background noise (including that from the longer scattering tails associated with lower frequencies, presumably caused by multitudes of other lenses).

We can be verify the results by looking at $P^{\prime\prime}=d^2P(\xi)/d\xi^2$ in Fig.~\ref{fig:derivatives} and using it to evaluate the magnification via $\mu=(1+fP^{\prime\prime})^{-1}$ (Eq.~\ref{eqn:magnification}).
For the incoming and outgoing echoes (positive and negative $\xi$, respectively), the extrema are $P^{\prime\prime}=(0.86, 0.27)$, so about mid-way through the track, one expects magnifications of about $(0.04, 0.12)$.
Relative to the main image, it is about the same for the incoming echo, since there the line of sight to the pulsar is barely affected by the lens.
But for the outgoing echo, the ratio almost doubles, since the lens demagnifies the main image by almost a factor two.
The magnification's dependence on wavelength enters via $f$; one thus expects $\mu\propto\lambda^{-2}$ as long as $P^{\prime\prime}f\gg1$, though in the ratio shown in Fig.~\ref{fig:magnification}, this is obscured for the outgoing echo by the fact that with decreasing wavelength the pulsar image also becomes fainter (in slightly less obvious ways, since it has $P^{\prime\prime}f\simeq1$), so the ratio increases faster than $\lambda^{-2}$ for that echo.
For the faint image during the transit, $P^{\prime\prime}$ reaches a minimum of $-1.37$ and thus one expects $\mu=-0.03$, which again should vary with~$\lambda^{-2}$.

While not all expectations are easy to test, in general the observations seem inconsistent with the predicted magnifications (see Fig.~\ref{fig:nov2021}).
In particular, while caustics could be missed as our observations consist of just 15 minute long on-beam recordings spaced apart by a full sidereal day, we do not see anything looking at finer frequency.
More generally, though both our and previous observations suggest that the echoes become brighter as they approach the main image, there has been no evidence of brightening at larger delays close to the start or end of the echo event, neither in the example shown in Fig.~\ref{fig:nov2021}, nor in other echoes we have observed or in the very strong 1997 event \citep{smith00, backer00, lyne01}.
However, about halfway down the incoming echo, the relative brightness of the echo is $\sim\!5\%$ for November 12 at $600 {\rm\, MHz}$, or $\sim\!4\%$ for November 9 at $450 {\rm\, MHz}$, which is comparable to what is expected.

Similarly, while it is difficult to measure relative brightnesses of overall profiles, given that these contain a sum of giant pulses with a large distribution of intensities, it is clear that the emission observed in the low magnification gap is more than an order of magnitude brighter than the predicted $\sim\!3\%$: we find the demagnification ranges from about 37\% at $800{\rm\,MHz}$ to 52\% at $400{\rm\,MHz}$.

We discuss two possible solutions to these mismatches in Section~\ref{sec:extensions}, suggesting that with a less smooth lens one might not have the caustics and brightening (Sect.~\ref{sec:noisy_skin}), while further lensing in the nebula might help fill in the gap (Sect.~\ref{sec:further_scattering}).

\subsection{Dispersion Measure Variations}\label{sec:dm_changes}

A direct test of our model would be through the variation of DM with time.
To do this properly would rely on detailed modeling, since the DM is different for different echo components (which do not necessarily trace the DM in the same direction; see Fig.~\ref{fig:summary}).
We have not attempted this analysis, but a qualitative comparison is possible using the approximate estimates of the DM made while detecting the pulses, which are determined by \review{minimizing the rise time of the frequency averaged pulse (see \citealt{serafinnadeau26}).}

\begin{figure}
  \centering
  \includegraphics[width=\columnwidth]{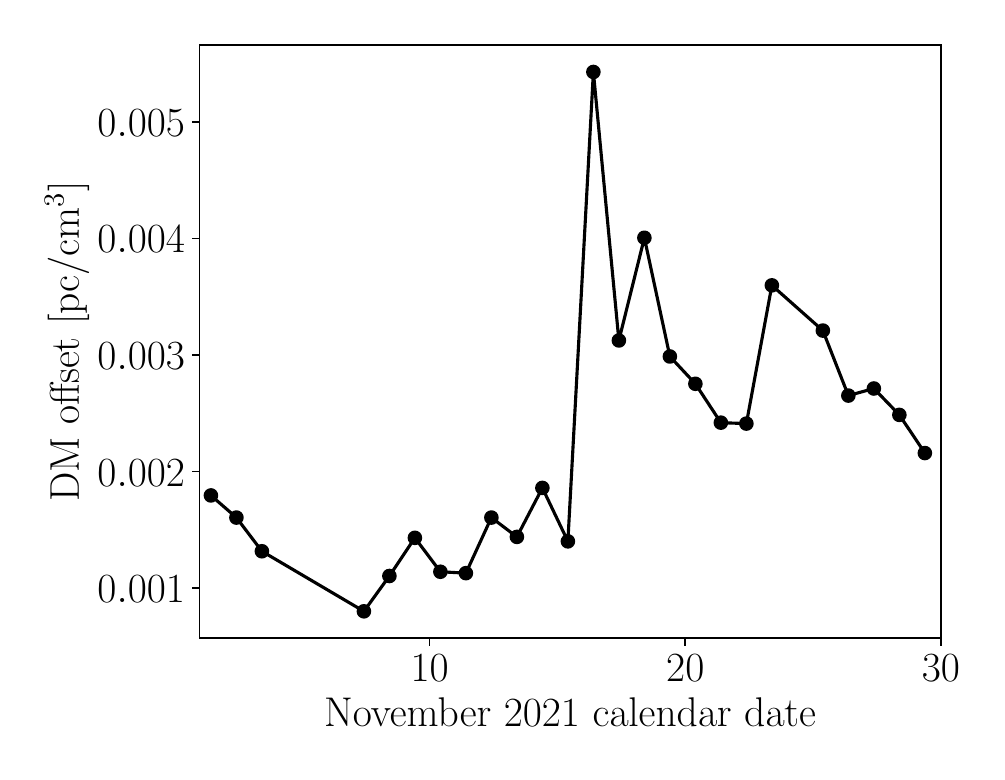}
  \caption{
    Measured DM offset through the month of November 2021, during which the echo shown in Figure~\ref{fig:nov2021} occurs.
    The offset is relative to $56.73{\rm\,pc/cm^3}$\review{, a typical DM during the observations from which the data in this paper were sourced \citep{serafinnadeau26}.}
    \label{fig:dm_data}
  }
\end{figure}

The resulting ``profile average'' DM measurements are shown in  Fig.~\ref{fig:dm_data}, and one sees that the overall shape is roughly consistent with that predicted.
First, as long as the main image is present (see the orange curve in Fig.~\ref{fig:summary}), there is almost no change in DM, because the main image dominates over the echo (cyan curve) in the pulse profile, and unlike the echo, the path does not go through any high density region.
Then, there is a big jump, as expected if only the faint image remained visible (green curve).
The jump is followed immediately by a drop, when the source brightens again because the other images come back (presumably dominated by the main image; blue curve), leveling off at an increased DM since it is now seen through larger column density.
Quantitatively, however, the numbers do not match: our model predicts a maximum increase in DM of about $8\times10^{-3}{\rm\,pc/cm^3}$, about twice the $\sim\!4\times10^{-3}{\rm\,pc/cm^3}$ observed.
Similarly, the predicted final offset is about twice that observed.

Of course, our value of ${\rm DM}_{\rm scl}$ was just an estimate, based on typical values inferred from a completely different set of pulses, so it may simply be that the true value is about half that.
If so, then we would need to decrease our width $T$ correspondingly in order to keep the maximum delays and durations consistent (as these depend on the maximum bending angles, which scale as ${\rm DM}_{\rm scl}/(T/2)$).
With that, the inner appearance and disappearance points would shift inward a little (the increase in bending angles more than compensated by the images shifting in), decreasing the duration of the shadow from $2.4$ to $2.0{\rm\,d}$, an amount that we cannot really constrain with our observations but is arguably slightly more consistent with what is seen in Fig.~\ref{fig:nov2021}.
The changes would also imply an increase in lens strength by the same factor, and thus the images would become fainter, but this would obviously not solve the broad inconsistencies in the magnification patterns found above.
Hence, like for the default model, one may have to appeal to having modulations on top of the smooth lens profile.

\section{Limitations and Extensions}\label{sec:extensions}

Our model is very simple and here we discuss its limitations as well as possible extensions.
We start at large scales, considering more general orientations of the filaments.
This does not change the shape, which will remain set by $P(\xi)$, but does affect the scales, via changes in ${\rm DM}_{\rm scl}$ and $T$.
Next, we consider the assumed density profile, which does affect the shape.
Then, we qualitatively discuss an extension beyond our model, where the large-scale structure is not smooth, so that, effectively, a combination of lenses contributes.
And finally we consider the influence of further lensing, by structures elsewhere in the nebula.

\subsection{Filament Orientation}\label{subsec:ellipse}

Our model assumed a cylindrical filament, oriented perpendicular to both the line of sight and to the pulsar motion, so that the cross-section was circular (see Eq.~\ref{eqn:circle}).
To generalize this to arbitrary orientation, we take the cylinder axis to have an inclination $i$ relative to the line of sight ($z$ axis), and a position angle $\Omega$ relative to the direction of pulsar motion on the sky ($x$ axis), i.e., we start with the cylinder around the line of sight, rotate it by $i$ around $y$, and then by $\Omega$ around $z$, giving a transformation,
\begin{equation}
  \begin{pmatrix}x^\prime\\y^\prime\\z^\prime\end{pmatrix}
  = \begin{pmatrix}
    \cos\Omega\cos i& - \sin\Omega& \cos\Omega\sin i\\
    \sin\Omega\cos i& \cos\Omega& \sin\Omega\sin i\\
    -\sin i& 0 &\cos i
  \end{pmatrix}
  \begin{pmatrix}x\\y\\z\end{pmatrix}.
\end{equation}
Here, given the symmetries, we can take $0^\circ\leq i\leq90^\circ$ and $0^\circ\leq\Omega<90^\circ$ without loss of generality, and hence assume positive $\sin$ and $\cos$ terms.

The points of interest are on the intersection of the cylinder with the plane defined by the line of sight and the pulsar motion, i.e., they should have $y^\prime=0$ and in the unprimed coordinates be constrained by $x^2+y^2=R^2$. Using $y^\prime=0$ to eliminate $z$ and defining an azimuth $\varphi$ such that $x=-R\sin\varphi$ and $y=-R\cos\varphi$ (which gives nicer signs), we find that the coordinates $x_s, z_s$ of the cross-section with the filament are given by,
\begin{align}
  x_s &= R\left.\frac{\cos\varphi}{\sin\Omega}\right.\\
  z_s &= R\left(\frac{\sin\varphi}{\sin i}
        +\frac{\cos\varphi}{\tan\Omega \tan i}\right).
\end{align}
One sees that this reduces to a circle traced by $\varphi$ for our default case, which corresponds to $i=90^\circ$ and $\Omega=90^\circ$.

Lensing will occur at the edge of the filament, at extrema in $x_s$, i.e., where $\varphi=0^\circ$ or $180^\circ$.
This implies an offset $x_{\rm edge}$ where the line of sight to the pulsar crosses the edge, projected filament skin width $T_{\rm edge}$, and radius of curvature $R_{\rm curv}$,
\begin{align}
  x_{\rm edge} &= \pm\frac{R}{\sin\Omega},\\
  T_{\rm edge} &= \frac{T}{\sin\Omega},\\
  R_{\rm curv} &= \frac{R\sin\Omega}{\sin^2 i},
\end{align}
where we used $R_{\rm curv}=(d^2x_s/dz_s^2)_{\rm edge}^{-1}$, with the derivative found by expanding $x_s$ and $z_s$ around $\varphi=0^\circ$.

Since a given echo only allows one to constrain ${\rm DM}_{\rm scl}=2n_{e,c}\sqrt{R_{\rm curv}T_{\rm edge}}$ and $T_{\rm edge}$, one obtains at best an estimate of the radius of the filament.
However, the inferred $T_{\rm edge}$ sets an upper limit to the true thickness $T$ of the filament's skin.

Finally, we note that in \citet{serafinnadeau24}, we used an angle $\psi$ between the proper motion of the pulsar and the normal to a sheet-like structure.
On the sky, this structure would trace the side of the cylinder, i.e., be normal to the cylinder axis, as projected on the sky.
Hence, one has $\psi=90^\circ-\Omega$, i.e., $\cos\psi=\sin\Omega$.

\subsection{Ionization profile}\label{subsec:profile}

\begin{figure*}
    \centering
    \includegraphics[width=1\textwidth]{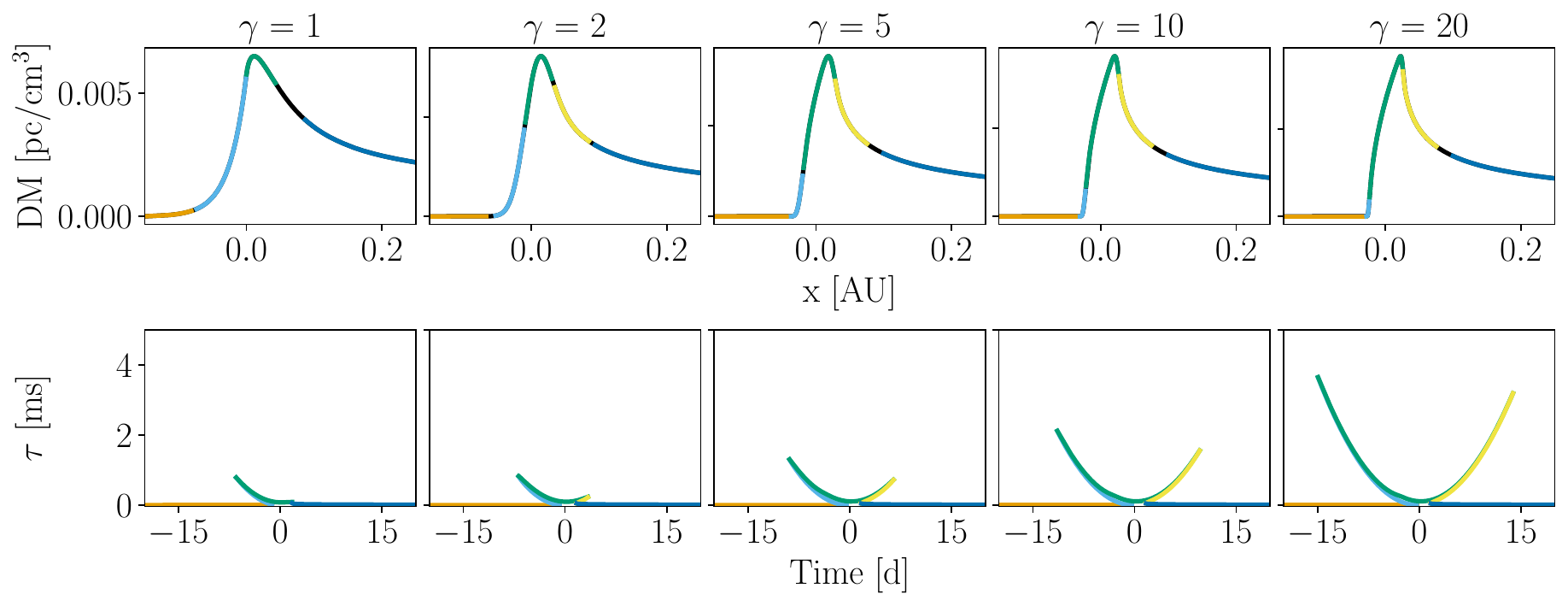}
    \caption{
        Simulated image delays at 600\rm{MHz} for a range of filament skin ionization profiles, with $\gamma=1$ corresponding to a Laplacian profile, $\gamma=2$ corresponding to a Gaussian profile, and larger $\gamma$ increasingly approximating a tophat ionization profile.
    }\label{fig:profile}
\end{figure*}

So far, we have assumed that the free electron density profile of the filament skin can be represented by a simple Gaussian.
However, reality is likely more complex.
To get a sense of the effects of different choices, we try a generalised Gaussian profile,
\begin{equation}
    \label{eqn:density_profile_mod}
    \Delta n_e(\rho) = n_{e,c}\frac{\gamma}{\Gamma(1/\gamma)} e^{-(2|\rho|/T)^{\gamma}}
\end{equation}
where $\gamma$ is a shape parameter (with $\gamma=2$ corresponding to a Gaussian and the profile approaching a top-hat function as $\gamma$ approaches infinity), and where, like with Eq.~\ref{eqn:density_profile}, the profile is normalised such that the column density across the skin is $n_{e,c} T$.
Inserting this in the column density integral (Eq.~\ref{eqn:dispersion}), one recovers that $\Delta {\rm DM}(x) = {\rm DM}_{\rm scl}P(\xi)$ (Eq.~\ref{eqn:scale_and_profile}), but with an adjusted profile,
\begin{equation}
    P(\xi)=\frac{\gamma}{\Gamma(1/\gamma)}\int_0^{\infty}e^{-(\varrho+\xi)^\gamma}/\sqrt\varrho\,d\varrho
\end{equation}

In Figure \ref{fig:profile}, we show DM profiles and implied image delays for a few choices of~$\gamma$.
One sees fairly drastic changes in the simulated echoes, with the durations increasing towards higher $\gamma$ due to the steeper column density gradients, and with the asymmetry decreasing as the parts of the profile that matter become more similar when the pulsar is inside or outside the shadow of the cylinder.
Conversely, towards lower $\gamma$ the durations decrease and the asymmetry increases, with the short echo all but disappearing at $\gamma=1$.

From these experiments, we conclude that a slight change in the density profile could easily account for the small discrepancies of our standard model with the observations in the asymmetry between the incoming and outgoing echoes.
Unfortunately, the sensitivity of the results to the precise density profile also make it difficult to infer reliable values of the model parameters, at least unless one can determine reliably also the change in DM during the echoes.

In addition, the interior of the filament is not, in principle, required to be fully neutral.
For a filament interior with a constant electron density \review{excess} $n_{e, i}$, the \review{excess} electron column density through the center of the filament is $2(n_{e, c}T + n_{e, i}R)$.
As the interior density increases, the column density gradient along the inner edge of the surrounding skin decreases, resulting in more asymmetric echo pairs.

When the electron column density through the filament interior surpasses that through the curved skin, echoes can no longer be produced in close pairs.
This establishes a strong upper limit on the ionization of the filament's interior for such echoes, when the column density surpasses the column density:
\begin{eqnarray}
	2(n_{e, c}T + n_{e, i}R) \lesssim {\rm DM}_{\rm scl} \nonumber\\
	\frac{n_{e,i}}{n_{e,c}} \lesssim \sqrt{\frac{T}{R}} - \frac{T}{R}
\end{eqnarray}
For a skin thickness $T \sim 0.05{\rm\, au}$, a filament radius $R \sim 10{\rm\,au}$, and a characteristic electron density in the skin $n_{e,c} \sim 10^3{\rm\,cm^{-3}}$,  the electron density within the filament is constrained to be $n_{e,i} \lesssim 70 {\rm\, cm^{-3}}$.

\subsection{Rough Skin}\label{sec:noisy_skin}

The model of lensing by a smooth ionized skin fails to provide a good description of the brightness of the echoes in the November 2021 event.
First, the predicted brightening near appearance and disappearance points is not seen.
This is unlike evidence for amplification near caustics  seen in extreme scattering events \citep{fiedler87, clegg98}, but similar to what is the case in previously observed echoes of the Crab, in none of which there have been signs of brightening of echoes at large delays.
Second, the magnification of the image left in the gap between arcs is predicted to be an order of magnitude fainter than is observed.

It is perhaps not surprising that our model does not match the observations in detail.
Indeed, it may be unrealistic to expect a filament to have an ionized interface that is smooth over an aspect ratio of $Z/T\simeq50$, especially in the context of the rather dynamic environment of the Crab nebula.
A perhaps more reasonable expectation is for the skin to have many smaller-scale perturbations in both width and distance from the center of the filament.
If so, instead of a single lens producing only a few point-like images one would have, effectively, multiple lenses forming many unresolved images, which would add together incoherently to form the observed, scattered images.

Similar multi-scale lensing has been appealed to in modeling lensing by outflows in binaries \citep{fangxi21, fangxi23}, and in this regime the overall form of the lensing structure can dominate the geometric evolution of the images, while the magnification behavior becomes dominated by the small scale variations \citep{jowthesis}.

One consequence of this will be that the caustics produced by the small scale lenslets become too small to be seen.
Furthermore, as these small-scale lenses likely have different strengths,
at larger scattering angles, fewer would contribute, suppressing the overall brightness, while closer to the line of sight, more would contribute, increasing it.
Indeed, it may even be possible for the main image to continue being visible as the pulsar passes behind the filament skin, merely becoming demagnified as its light is partially scattered away from the line of sight.
For our event, however, this seems unlikely, as then one would expect the width of the observed profile to remain roughly constant through the gap, unlike what is seen.

If the lens is indeed effectively composed of multiple smaller ones, the inferred properties of the overall structure will no longer match the physical ones.
In particular, the smaller lensing structures will necessarily have smaller ${\rm DM}_{\rm scl}$, and must thus have smaller physical thickness to ensure the maximum scattering angle remains the same.
This also means that the model value of $T$ and observed features like the width of the gap between echo arcs do not reflect the physical thickness of the skin but instead the amplitude of the perturbations of the lens locations.
Furthermore, the column density plateau through the center of the filament, which should roughly be $2Tn_{e,c}$, will be lower (given that $n_{e,c}$ is constrained by the optical line emission).

There may be some evidence for small-scale lensing in the much stronger 1997 event.
In particular, in Fig.~1 of \cite{backer00}, which shows its evolution at $327{\rm\,MHz}$, one sees that when the pulsar reappears, its DM is not only larger, as expected if our line of sight now crosses a lensing structure, but the scattering time has also increased substantially, suggesting the structure is not smooth.

\subsection{Further Scattering}
\label{sec:further_scattering}

So far, we have treated our lens in isolation, but the fact that the pulsar emission always has a scattering tail shows that the pulsar's radiation is affected by further structures, likely partially in the Crab nebula itself and partially in the interstellar medium.

The scattering regions further in the Crab nebula may affect what we
see, since those see the lens from slightly different perspectives.
For instance, \review{using that across our observation band,} the broader scattering tail \review{is confined within}
$\tau\simeq0.1{\rm\,ms}$ \review{(see Fig.~\ref{fig:nov2021}), and assuming this further scattering} is due to matter in the middle of the nebular material, at $d_{ps}\simeq1{\rm\,pc}$, then this implies angles differing by
$\delta=(2c\tau/d_{ps})^{1/2}\simeq0.3{\rm\,arcsec}$.
Hence, when the strong images disappear from the direct line of sight, they will still contribute via rays that are scattered, for about $d_{\rm ps}\delta / v_{\rm eff}\simeq2{\rm\,d}$, filling in the gap.

The 1997 event may again present some evidence for this effect: in Fig.~1 of \cite{backer00}, one sees that while the pulsar is faint, two components are visible, one each with the old and new DMs, suggesting we are seeing scattered bits of the brighter profiles on both sides.
Indeed, the old image seems to get narrower and slightly delayed as it fades, as would be the case if it was seen through increasingly large scattering angles on a second screen.

\review{In detail, one needs to take into account that scattering is highly frequency dependent, scaling roughly with $\nu^{-4}$.
Therefore, at higher frequencies, it may be more difficult to fill in gaps.
On the other hand, gaps are smaller at higher frequencies, and magnifications larger (see Fig.~\ref{fig:magnification}).}

\section{Ramifications}\label{sec:ramifications}
In this paper, we have shown that the basic properties of echoes produced in the Crab nebula can be reproduced in a model where the pulsar passes behind filaments with thin ionized surfaces.
In this picture, the column density profile will be asymmetric, with a much steeper gradient on the outside than on the inside, leading naturally to echoes that are nearly one-sided, as was observed for the 1992 and 1997 events first presented by \citet{smith00}, as well well as in the various echoes shown in \citet{serafinnadeau24}.

We create a simplified model, in which we take the filament to be cylindrical and the surface to have a Gaussian electron density profile.
We compared our simulations with an echo from November 2021 observed with CHIME, which is simple and isolated from other events, and find that it reproduces the time evolution in detail.
However, it only reproduces the magnifications qualitatively: the simulations exhibit caustics which are not observed in actual data, and predict far too much demagnification during the gap between echo arcs.
We think the likely cause is that the skin has more structure than we have assumed, causing multiple images.
It would be worth exploring this in more detail.

Our model makes a number of general predictions.
First, events with longer incoming arcs will be the result of the Crab entering the filament's shadow, and should thus correspond to increases in DM across the event, while receding long arcs indicate the pulsar is leaving the filament behind and should correspond to decreases in DM.
We also expect filaments to produce pairs of events: one as the source enters its shadow and another when it leaves.
In principle, in our idealized case, the events should be identical except for being flipped.
Of course, in reality they may not be that similar, but long-term tracking of echoes should find that the number of incoming and outgoing arcs evens out.

Second, since filaments have finite size, there should be some where the pulsar only gets close to but never crosses behind the filament, i.e., there should be some echoes that approach zero delay but never reach it, instead again receding to larger delays.
The statistics of such echoes seen in a larger survey should give a good constraint on the typical size of filaments.
Models for them would depend on the details of the geometry at the filament tip.
It may be possible to describe this with simple models, such as domed filament ends, but it may also give direct evidence for, e.g., the larger globular tips described in \cite{hester96}, which are extended as a result of cooling.

Third, taking the curvature radii of $\sim\!10{\rm\,au}$ at face value, the apparent filament diameters would be of order $10{\rm\,mas}$, a scale crossed by the Crab pulsar in about one year.
The filaments emit line emission, and while it would be difficult to resolve them with current instrumentation, it may be possible to identify individual filaments responsible for (clusters of) echoes, especially if one uses velocity information (e.g., \citealt{martin21}).

\review{While we describe the results above in terms of our hypothesis of a small filament with an ionized skin, we stress again that the basic requirement to reproduce the observed echo properties is to have thin sheets of dense ionized material seen nearly edge-on.
  Indeed, any structure that produces column density profiles qualitatively similar to those described in this paper would also explain the properties of echoes; the question would just be whether or not these are related to the larger structures observed in the nebula.}

\review{Arguably, the biggest question posed by the echoes is how it is possible to obtain structures sufficiently thin to produce the required large electron column density gradients, given that modeling of the nebula suggests ionization depths that are much larger, $\sim 10^{3} {\rm \, au}$ \citep{pequignot83, sankrit98, loll13, priestley17, das20}.
Some mechanism to confine matter, such as with magnetic fields, may be required.}

Finally, we note that in our model we followed the formalism of \cite{simard18}, since  the electron columns density near ingress and egress have essentially the same form as that of the parabolic sheet assumed by those authors.
Our results thus provide supporting evidence for the existence of such structures.
But they also show that the assumptions of a single smooth structure may be too limiting, and that this may affect the predicted magnifications in particular, similar to what was noted by \cite{fangxi23} and \cite{jow24}.

\vspace{5mm}
\textit{Acknowledgements}:  We thank Jing Luo, Fang Xi Lin, and Dylan Jow for help during the various stages of this project, the Toronto scintillometry group generally for useful discussions, and Chris Thompson for pointing out the possibility of constraining the electron density inside the filament.
\review{We also thank the anonymous referee for their detailed review, which helped make the paper much clearer.}
M.Hv.K. is supported by the Natural Sciences and Engineering Research Council of Canada (NSERC) via discovery and accelerator grants, and by a Killam Fellowship.
Computations were performed on the Sunnyvale computer at the Canadian Institute for Theoretical Astrophysics (CITA).

\vspace{5mm}
\facilities{CHIME \citep{chime}.}

\vspace{5mm}
\software{astropy \citep{astropy13, astropy18, astropy22}, numpy \citep{numpy20}, matplotlib \citep{matplotlib07}, scipy \citep{scipy20}.}

\appendix

\section{Semi-analytical Solution for the Column Density}
\label{sec:appendix}

For the case of a Gaussian density distribution given by Eq.~\ref{eqn:density_profile}, the column density integral can be described by a parabolic cylinder function to high precision.
To see this, we start with the second equality in Eq.~\ref{eqn:dispersion}, normalizing the DM to that at the middle of the cylinder ($x=0$) by dividing by $2Tn_{e,c}$, and change variables to dimensionless ones relative to the tangent point of the sheet, $\hat\rho=2(\rho-R)/T$ and $\hat{x}=2(|x|-R)/T$,
\begin{align}
  \frac{\Delta {\rm DM}(x)}{2Tn_{e,c}}
  &= \frac{2}{T\sqrt\pi} \int_{|x|}^\infty
  \frac{e^{-(2|\rho-R|/T)^2}}
       {\sqrt{1-(x/\rho)^2}}d\rho\nonumber\\
  &= \frac{1}{\sqrt\pi} \int_{\hat{x}}^{\infty}
  \frac{e^{-\hat\rho^2}}
  {\sqrt{1-\left(\frac{1+\hat{x}T/2R}{1 + \hat\rho T/2R}\right)^2}}d\hat\rho.
  \label{eqn:dm_dimensionless}
\end{align}
Next, we use that $\hat\rho T/R\ll 1$ to approximate $(1 + \hat\rho T/2R)^{-2}\simeq1-\hat\rho T/R$,
\begin{align}
  \frac{\Delta {\rm DM}(x)}{2Tn_{e,c}}
  &\simeq \frac{1}{\sqrt\pi} \int_{\hat{x}}^{\infty}
           \frac{e^{-\hat\rho^2}}
           {\sqrt{\hat\rho\frac{T}{R}-\hat{x}\frac{T}{R}-(\hat{x}\frac{T}{2R})^2}}d\hat\rho\nonumber\\
  &= \sqrt{\frac{R}{T}}\frac{1}{\sqrt\pi} \int_{\hat{x}}^{\infty}
      \frac{e^{-\hat\rho^2}}
      {\sqrt{\hat\rho-\xi}}d\hat\rho,
  \label{eqn:dm_approx}
\end{align}
where in the last equality we introduced $\xi\equiv \hat{x}(1+\hat{x}T/4R)$.
One sees that close to the tangent point of the sheet, $\xi\simeq\hat{x}$, but it starts to deviate further away.
Expressing it in terms of $x$, one finds $\xi=-(R/T)(1-(|x|/R)^2)$.
Here, with our choice of signs it is clear that for $|x|\le R$, one can write $\xi=-(R/T)\cos^2i(x)$, where $i(x)$ is the inclination of the sheet at $x$ (with $i=90^\circ$ edge-on, at the tangent point).
As a result, for $\hat{x}\ll0$, where $-\xi\gg\hat\rho$ for the relevant range, one recovers that the right-hand side equals $1/\cos i(x)$.

Finally, for the lower limit of the integral, we can safely replace $\hat{x}$ with $\xi$, since the quadratic term in $\xi$ is important only when the peak of the integrand is far away from $\hat{x}$.
Doing that, and shifting the integration domain to $\varrho=\hat\rho - \xi$, we find that the integral can be expressed in terms of a parabolic cylinder function, $U(a, z)$ with $a=0$ \citep[\href{https://dlmf.nist.gov/12.5.E1}{\S12.5, Eq.~1}]{NIST:DLMF},
\begin{equation}
 P(\xi) = \frac{1}{\sqrt\pi}\int_{0}^{\infty} \frac{e^{-(\varrho+\xi)^2}}{\sqrt{\varrho}}d\varrho
 = \frac{e^{-\frac12\xi^2}}{2^{1/4}}U(0, \sqrt2\xi).
  \label{eqn:dm_profile}
\end{equation}
This also allows one to calculate derivatives easily via recurrence relations \citep[\href{https://dlmf.nist.gov/12.8.E10}{\S12.8, Eq.~10}]{NIST:DLMF},
\begin{equation}
  \frac{d^m}{d\xi^m} e^{-\frac12\xi^2} U(0, \sqrt2\xi) = (-\sqrt2)^m e^{-\frac12\xi^2}U(-m, \sqrt2\xi).
  \label{eqn:derivatives}
\end{equation}
For derivatives with respect to $x$, one can use that $d\xi/dx=(2/T)(1+(|x|-R)/R)$ and $d^2\xi/dx^2=\sign(x)2/RT$.
Near the peak, where $|x|\simeq R$ and $\xi\simeq2(|x|-R)/T$, one has $d\xi/dx\simeq\sign(x)2/T$.

\begin{figure}
  \centering
  \includegraphics[width=\hsize]{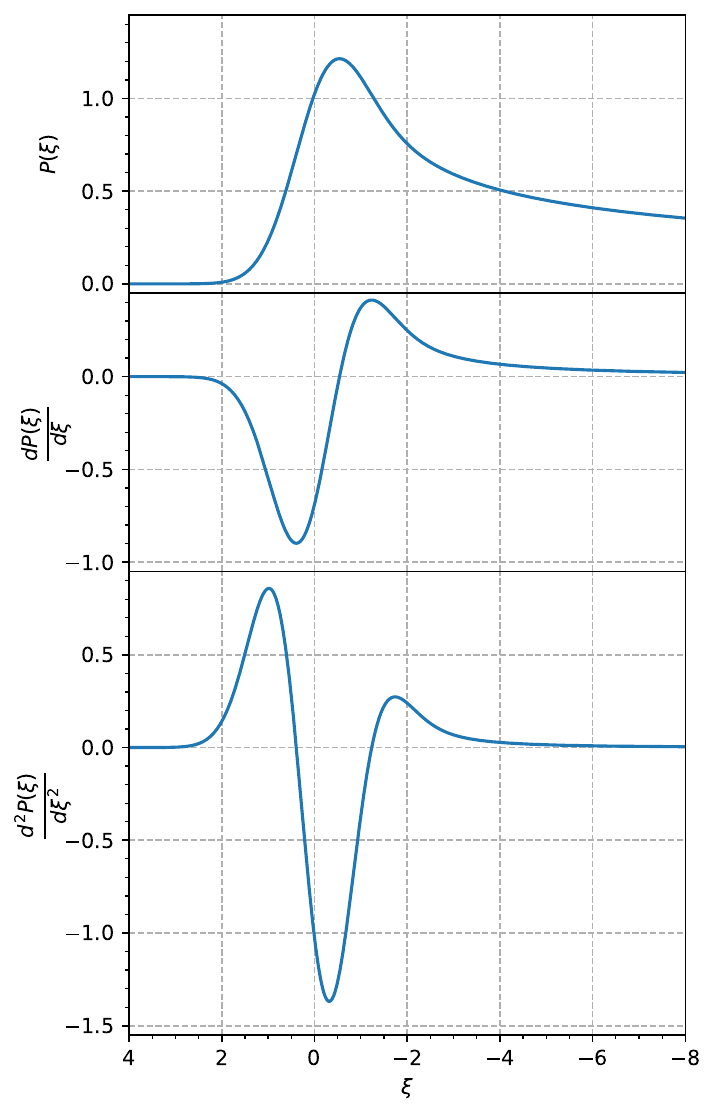}
  \caption{
    The normalized lens profile and its first and second derivatives.
    Images are bent proportionally to $dP/d\xi$, and pairs of them (dis)appear when $d^2P(\xi)/d\xi^2$ equals a number that depends on the lens strength (see Sect.~\ref{sec:delays}).
    The dimensionless coordinate $\xi$ is chosen to run from positive to negative (from outside to inside the filament) to match the 2021 November echo discussed in this paper.
    \label{fig:derivatives}
  }
\end{figure}

\begin{deluxetable}{rrrrrr}
\tablecaption{\label{tab:extrema}Roots and Extrema of $P(\xi)$ and its Derivatives.}
\tablehead{
  \colhead{$d^3P/d\xi^3$}&
  \multicolumn{2}{c}{\dotfill$d^2P/d\xi^2$\dotfill}&
  \multicolumn{2}{c}{\dotfill$dP/d\xi$\dotfill}&
  \colhead{$P$}\\
  \colhead{roots}&
  \colhead{extr.}& \colhead{roots}&
  \colhead{extr.}& \colhead{root}&
  \colhead{extr.}
  }
\startdata
$-1.7396$ & $ 0.2734$ & $-1.2348$ & $ 0.4129$ & $-0.5409$ & $ 1.2143$\\
$-0.3194$ & $-1.3689$ & $ 0.3895$ & $-0.8985$ & \nodata   & \nodata  \\
$ 0.9810$ & $ 0.8583$ & \nodata   & \nodata   & \nodata   & \nodata
\enddata
\end{deluxetable}

We show $P(\xi)$ and its first two derivatives in Figure~\ref{fig:derivatives}, and give roots and extrema in Table~\ref{tab:extrema}.
These can be used to find where the outer pairs of images will appear: for a strong lens, one will be close to a root of $d^2P(\xi)/d\xi^2$ and hence $dP(\xi)/d\xi$ will be close to an extremum.
For the inner pairs of images, one can estimate locations using expansions of $P(\xi)$ and its derivatives for large $|\xi|$,
\begin{equation}
  \frac{d^mP(\xi)}{d\xi^m} \sim \left\{
    \begin{array}{l}
      (\frac12)_m\xi^{-\frac12-m}\hfill(\xi\ll0),\\
      (-1)^me^{-\xi^2}(2\xi)^{-\frac12+m}\quad(\xi\gg0),
    \end{array}\right.
\end{equation}
where $(\frac12)_m=\frac12\times(\frac12+1)...(\frac12+m-1)$.
For given second derivative $P^{\prime\prime}_g$, for negative $\xi$, one will have $\xi_{-}\simeq(\frac43 P^{\prime\prime}_g)^{-2/5}$ and infer $P^\prime_{-}\simeq\frac12\xi_{-}^{-3/2}$;
for positive $\xi$, with some trial and error, $\xi_{+}\simeq(-\log P^{\prime\prime}_g)^{1/2}+2^{-3/2}$ and $P^{\prime}_{+}\simeq-e^{-\xi_{+}^2}(2\xi_{+})^{1/2}$.
A power-law expansion around the inferred values, $P^{\prime}=P^{\prime}_{\pm}(P^{\prime\prime}/P^{\prime\prime}_g)^\alpha$, will have powers $\alpha_{-}=3/5$ and $\alpha_+=1+[2\xi_+(\xi_+-2^{-3/2})]^{-1}$ for negative and positive $\xi$, respectively.

For numerical evaluation, one can use the {\tt pbdv} function in {\tt scipy} \citep{scipy20}), though note that its arguments are like those for an older notation for a parabolic cylinder function, $D_\nu(z)$, with $\nu=-a-\frac12$.
A numerical problem with this is that for large $\xi$, the function grows exponentially (compensated by the $e^{-\xi^2/2}$ term in Eq.~\ref{eqn:dm_profile}).
To avoid dealing with this, one can rewrite $P(\xi)$ in terms of
hypergeometric functions,
\begin{align}
  P(\xi)
  &= \frac{1}{\sqrt\pi}
    \left[\tfrac12\Gamma(\tfrac14)M(\tfrac14, \tfrac12, -\xi^2)
    - \xi\Gamma(\tfrac34)M(\tfrac34,\tfrac32,-\xi^2)\right],
  \label{eqn:profile_hyp}
\end{align}
where $M(a, b, z)$ is Kummer's confluent hypergeometric function of the first kind (also written as $\Phi(a, b, z)$ and equal to the generalized hypergeometric series ${}_1F_1(a; b; z)$, implemented in {\tt scipy} as {\tt hyp1f1}).

\vfill

\bibliography{main}{}
\bibliographystyle{aasjournal}

\end{document}